\documentclass[
]{ceurart}

\usepackage{listings}
\begin{document}

\copyrightyear{}
\copyrightclause{}


\title{Mapping Climate Change Research via Open Repositories \& AI: advantages and limitations for an evidence--based R\&D policy--making.}


\author[1]{Nicandro Bovenzi}[%
orcid=0000-0002-7764-3974,
]
\address[1]{SIRIS Lab, Research Division of SIRIS Academic, 08003, Barcelona, Spain}

\author[1]{Nicolau Duran-Silva}[%
orcid=0000-0001-5170-4129,
]

\author[1]{Francesco Alessandro Massucci}[%
orcid=0000-0001-5405-8759,
email=francesco.massucci@sirisacademic.com,
]
\cormark[1]

\author[1]{Francesco Multari}[%
orcid=0000-0001-8509-3178,
]

\author[1]{C\'esar Parra--Rojas}[%
orcid=0000-0003-3625-9412,
]

\author[1]{Josep Pujol-Llatse}[%
]
\cortext[1]{Corresponding author.}

\begin{abstract}
In the last few years, several initiatives have been starting to offer access to research outputs data and metadata in an open fashion. The platforms developed by those initiatives are opening up scientific production to the wider public and they can be an invaluable asset for evidence--based policy--making in Science, Technology and Innovation (STI). These resources can indeed facilitate knowledge discovery and help identify available R\&D assets and relevant actors within specific research niches of interest. Ideally, to gain a comprehensive view of entire STI ecosystems, the information provided by each of these resources should be combined and analysed accordingly. To ensure so, at least a certain degree of interoperability should be guaranteed across data sources, so that data could be better aggregated and complemented and that evidence provided towards policy--making is more complete and reliable. Here, we study whether this is the case for the case of mapping Climate Action research in the whole Denmark STI ecosystem, by using 4 popular open access STI data sources, namely OpenAire, Open Alex, CORDIS and Kohesio.
\end{abstract}


\maketitle
\vspace{-1cm}
\section{Introduction}
To inform their decisions, policy-makers in the Science, Technology and Innovation (STI) sector typically need ``maps'', either at a territorial or at an institutional level, to understand what is researched and by whom. Generally, those maps need to provide information about the research and innovation topics and about the relevant actors linked to them, so that effective policy-actions could be proposed, by covering the right scientific domains and by being catered for the adequate users. These maps need to be comprehensive, so to extensively cover {\em i.} the whole STI value chain (from basic research up to industrial innovation), {\em ii.} the different scientific domains and  {\em iii.} all possible relevant actors. As such, these maps should rely on different data sources that could offer the broadest possible view of STI inputs and outputs. 
For instance, they could be built on top of data covering scientific publications, R\&D projects funded by multiple parties (e.g., the region, the EU), patents, etc. One of the major challenges faced at a policy level is precisely that of identifying and combining information proceeding from different data sources, and to allow stakeholders to engage in turn with the processes analytical process.
Some major challenges faced at a policy level arise because many of those data sources are not openly available (undermining therefore the participatory processes), they are not interoperable in terms of data classification schemes and institutional identification (therefore limiting transversal analyses) and they are hardly manageable by non--expert users \cite{Fuster2020}.

These limitations become even more pressing when Science is called into action to tackle societal issues such as globalisation, the negative effects of new technologies, climate mitigation, 
and, more recently, the COVID-19 pandemic. This aspect is very relevant, because STI policy-makers are trying worldwide to mobilise Knowledge towards the solution of these outstanding problems. At the global level, the United Nations established the UN 2030 Agenda for Sustainable Development \cite{desa2016transforming} and formalised the 17 Sustainable Development Goals \cite{sdg2019sustainable} (hereafter, SDGs) to achieve such objectives. To guide policy efforts to steer science towards the SDGs, the UN also defined a series STI for SDGs Roadmaps \cite{sti4sdgs}. 
At the European level, the Commission is implementing the STI for SDGs roadmap by accelerating the green transition via the Green Deal \cite{19greendeal} and by allocating funds in the cohesion policy framework and the Horizon Europe programme to mobilise European research and innovation ecosystems towards the tackling of outstanding Societal Challenges.
However, because of the above difficulties, it remains fairly difficult to monitor how STI ecosystems are responding to these policy frameworks.

Fortunately enough, the first issue may be mitigated by the emergence of several initiatives that, spurred by the Open Science movement, are making STI input and output data (e.g., projects, publications, datasets) increasingly available in an open fashion \cite{hutchins2021tipping}. Projects like Crossref \cite{crossref}, OpenAire \cite{openaire}, OpenAlex \cite{openalex}, Open Citations \cite{peroni2020opencitations}, Open Abstracts \cite{i4oa} and ORCID \cite{orcid} are providing different, ample and complementary access to research outputs data and metadata that may greatly help overcoming the accessibility limitations of the analytical challenges at the STI policy level. But while data availability is slowly becoming less of an issue, the possibility of seamlessly carrying out analyses across those data sources has not been satisfactorily dealt with yet. However, encouraging signals emerge from the AI and tech industry: big tech players are changing the way in which they monetise data \cite{weber2022ai} and they are now sharing assets and technological outputs for the common good, generating a commonality that eases innovation \cite{chesbrough2006open}. Today, datasets and language  models created by the likes of Google \cite{devlin2019bert}, Facebook \cite{lewis2019bart} or the Allen Institute of AI \cite{scibert, cohan2020specter} 
are openly available to analyse
large STI datasets barely accessible up to some years ago. 

In this paper, we present a proof of concept of such an endeavour, by gathering data for the whole Danish STI ecosystem\footnote{We choose Denmark as our case study because (i) it is a medium-size country and the size of its scientific production is such that one can practically retrieve all the documents from publications repositories, (ii) Danish R\&D ecosystem is internationally visible and competitive and (iii) Denmark is internationally acknowledged as being one of the leading countries in terms of climate action policies and efforts (1st in the world according to the 2022 Environmental Performance Index (EPI)).} from 4 different data sources, namely:  
{\em i.} the CORDIS database\footnote{\tiny\url{https://cordis.europa.eu/}}, through the UNICS \cite{Gimenez2018UNiCSTO} platform, for H2020--funded R\&D projects, {\em ii.} the Kohesio linked--data portal\footnote{\tiny\url{https://kohesio.ec.europa.eu/}} for Regional R\&D projects funded by the EC Cohesion initiative, and the {\em iii.} OpenAlex \cite{openalex} and {\em iv.} OpenAIRE\cite{openaire} repositories for publications and other scientific outputs (these last being accessed through their respective APIs). After gathering these records, we use open knowledge--bases \cite{duran2019controlled} and text mining to showcasing how research in emerging fields (such as the SDGs) can be gathered from open data sources and identified by means of modern, openly available pre-trained language models. Finally, we demonstrate how gaps in taxonomic classifications across datasets may be filled by means of topic modelling and Deep Learning textual classifiers, by using the ERC panels as a paradigmatic example. They are trans--European Science classification scheme, which is (rightly) considered by policy--makers a convenient way to categories (and frame) excellent research.

In this way, the present paper aims therefore at tackling the following research questions:
\begin{itemize}
  \item Is it possible to use open repositories to inform policies in the field of STI?
  \item How do the various available solutions differ among them? (in terms of discipline and affiliation coverage)
  \item What is the availability of textual/disciplinary information?
  \item How can one obtain a reliable mapping of the available assets in a local STI ecosystem?
  \item Is it possible to identify pertinent actors?
  \item How does this apply to the specific case of Science applied to Climate issues?
  \end{itemize}


\section{Background}

\subsection{Mapping research concerning the Sustainable Development Goals (SDGs)}

Interpreting the potential breadth of each SDG and identifying which STI outputs ``contribute'' to them is a challenging task, because goals are broad and open to subjective interpretation as well subject to local declination \cite{rafols2021visualising}, and, additionally, STI records have complex domain-specific jargon, and they do not usually make reference to the application or possible social impact of the research activity. While the SDG targets and indicators \cite{united2018global} offer some objectivity to understand the scope of each goal \cite{bergen_sdgs}, it is still a challenge translating from policy language to the actual concepts used by researchers to talk about their studies on potential solutions having a positive impact on SDGs. Several projects have attempted to propose systems to link SDGs to the text of some STI output. Among the proposed methods, we can find from supervised machine-learning approaches trained on publications extracted from specific keywords \cite{hajikhani2021interrelation, amel2021nlp} to lists of keywords built by domain experts \cite{aurora_network2020search, confraria2021countries}.

Here, we identify SDG--related research by using a collection of SDG keywords (a controlled vocabulary) built by SIRIS Academic and openly available in Zenodo \cite{duran2019controlled}, based on a hybrid approach that use automatic methods for enriching human-crafted keywords.

\subsection{Topic Modelling}

Topic Modelling (TM) is an unsupervised classification problem in machine learning that aims at `discovering' the unknown topics linked with a specific collection of texts. This approach turns out to be an extremely useful tool to gain an emergent analysis of the research focus of institutions or thematic perimeters \cite{Fuster2020}, as for instance about climate change \cite{callaghan2020topography}, assisted reproduction \cite{garcia2020mapping}, and others \cite{liu2021tracing, griffiths2004finding}. While different methods and algorithms have been proposed to detect the topics, most of them share the idea that the topics provided are stacks of tightly related words that co-appear consistently in the observed texts.  As such, the topics yielded by the algorithm emerge from - and are characterised by - the language actually used in the texts at hand. Recently, the use of pre-trained language models (PLM) based on Transformers \cite{vaswani2017attention} such as BERT  (Bidirectional Encoder Representation from Transformers) \cite{devlin2019bert} is becoming increasingly popular for topic modelling \cite{topicbert1, berttopic3, bertopic4}. Models pre-trained in scientific literature, such as BioBERT \cite{biobert}, PubMedBERT \cite{pubmedbert}, or SciBERT \cite{scibert}, have demonstrated improvements compared with general-domain models in tasks  dealing with highly technical texts. 

Here, we perform Topic Modelling by clustering the vectorial representation of the analysed documents yielded by a BERT model pre-trained in scientific literature.

\section{Data sources}

We investigate the use of the following research and innovation data sources and the interoperability between them, filtering for publications produced by Danish institutions between 2014 and 2019:

\begin{itemize}
    \item {\bf OpenAlex} \cite{openalex} is a fully-open scientific knowledge graph, that contains around 209M works and 2013M disambiguated authors that are open under a CC0 license. 
    \item {\bf OpenAIRE} \cite{openaire} indexes 139 million publications. The most important objective of this platform is to offer all the EC founded initiatives the infrastructure the tools to become open accessible. 
    \item {\bf CORDIS} (Community Research and Development Information Service)\footnote{\url{https://cordis.europa.eu/
    }} contains data and metadata related to R\&D projects and related organisations which have received  funding by the European Commission under the H2020 and FP7 framework programs. 
    \item {\bf Kohesio}\footnote{\url{https://kohesio.ec.europa.eu/}} is an open database offering access to the R\&D projects funded by Cohesion Funds of the EU Commission in 2014-2020 programming period. It contains 1.5 million projects and approximately 500,000 beneficiaries. The data is distributed under a Creative Commons Attribution 4.0 International license.
\end{itemize}

\subsection{Data collection and selection}

\textbf{OpenAlex} --- Data can be accessed through API or by downloading a dump. For this article, the API method was used together with a filter for the Danish institutions and by filtering the date of publication between 2014 and 2019. For legal reasons, the abstracts do not appear as a text but as an inverted matrix, or as ``InvertedIndexAbstract", as explained in the documentation of the project. This encodes the words in the abstract and their positions. Plain texts from those matrices have been reconstructed so that they can be fed to our Topic Modeling tools.\\

\noindent\textbf{OpenAIRE} --- To download the data via API, we had to comply with the limitation on the maximum number of results per query set to 10,000 documents.  This was done by splitting the initial query in many sub-queries over reduced time ranges. The response was initially generated in JSON format, but it was not always possible to parse the JSON file correctly. Hence, the XML format was finally adopted.
The unreliability of certain metadata was also a limit to our analysis, because of missing abstracts or impossibility to track the language of the document (incorrect or missing information about the original language).
Since a publication may be collected by more than one source ({\em e.g.} several university internal repositories), it can have different sets of metadata.  For instance, the \textit{DateOfAcceptance} may refer to the date the document was uploaded to the (university) repository and hence be different for different repositories.
This makes very hard to compute trends over time (e.g, number of publications per year).\\

\noindent\textbf{CORDIS} --- The data are accessible through Open Data license \cite{cordis}  and they are monthly updated, afterwards provided in the format of CSV, XML and Linked Open data in the CORDIS website. In the case of this article,  we have collected the CORDIS records from UNiCS \cite{Gimenez2018UNiCSTO}, an open data platform based on semantic technologies for science and innovation policies developed by SIRIS Academic. Although project data is generally of good quality, the information regarding the localization of the project participant organizations is not always correct. This limitation was circumvented by the data cleaning operations carried out in order to integrate CORDIS data in UNICS platform.\\

\noindent\textbf{Kohesio} --- Many projects have a missing description (abstract) or an unsatisfactory one (the name of the project appears instead of the description, or just few words without a context). This issue limits the identification of Climate Change related projects through analyzing the textual content of the abstract.

\section{Materials and Methods} \label{sec:matMethods}

\subsection{SDG Mapping via controlled vocabularies}

SIRIS Academic developed a methodology \cite{duran2019controlled} based on machine learning and artificial intelligence to build an extensive SDG controlled vocabulary that can be exploited to map STI contributions to SDGs. The controlled vocabulary defines the semantic content of each SDG and related targets, and it can be used to classify textual records by detecting the presence of its key terms. 
The method for defining the vocabulary combines expert knowledge (namely, to define and limit the ``semantic'' breadth of each SDG) and on machine learning to rapidly scale the set of key terms defined by the domain experts. A first version of the vocabulary was developed by SIRIS in 2019, while in this work we use a new, revised version which features several additional terms with respect to the original vocabulary. We finally have carefully crafted matching rules that take into account permutations of words and that allow words within concept to be within a certain distance. 

\subsection{BERT-based topic modelling}
\label{sec:BERT}

To perform topic modelling on scientific documents, we use SPECTER \cite{cohan2020specter}, a BERT-based pre-trained language model \cite{devlin2019bert}. SPECTER is pre-trained on large corpora of scientific literature, by using citation information to reduce loss, as a strategy for including additional knowledge. We use SPECTER to encode STI in n-dimensional vectors and cluster them to find semantically similar text groups, by following the advice of its authors, who recommend its use without fine-tuning. Here, topics are found by means of the K-Means clustering algorithm on top of the encoded vectors. To find the best number of topics, we ran the K--means by varying the number of clusters and we computed the so-called \textit{Within Cluster Sum of Squares(WCSS)} and \textit{Distances between cluster centroids} (DBCC): we eventually chose to extract 30 clusters ({\em i.e.}, topics) by selecting the region where these two metrics did not fluctuate excessively and by qualitatively choosing the best trade off between the semantic ``richness'' of the topics and the overall number of topics (in order not to have neither too large topics nor too little ones).
Each cluster is therefore a topic and close vectors are thematically--related documents. Inspired by \cite{mei2007automatic, lau2011automatic}, topic labels have been finally obtained by ranking per frequency nouns and keywords of the documents belonging to each topic and by searching them in reports, policy documents and scientific literature: the labels are obtained by abstracting the titles of the respective documents.

\subsection{Weakly supervised text classification in the panels of the European Research Council (ERC)}

Traditional bibliometric taxonomies lack the necessary granularity and nuances to classify texts into finely defined domains of interest for STI policy--makers \cite{Fuster2020}. Additionally, bibliometric categories are not compatible across data sources, limiting the possibility of carrying out analyses across data sources. Here, we propose to go beyond such a limitation by using textual classifiers and by categorising all analysed textual information into the 25 European Research Council (ERC) Panels. By construction, the ERC covers all fields of science, by focusing primarily on pioneering and excellent research, that could imprint transformative changes in research landscape. The ERC is a pan--European science classification scheme, which is looked at as a framework for excellent research by STI policy--makers. Because it is independent from the bibliometric categories used in databases of scholarly communications and since it might be particularly relevant to also classify publications within this popular scheme, we decided to use it as a toy example for the case study we explore in our paper.

To train a multi-label text classification on ERC panels, we implement \textit{fine-tuning} as proposed by \cite{devlin2019bert}, on SPECTER \cite{cohan2020specter}. With fine-tuning, one adapts a pre--trained model to a downstream task: for text classification, this boils down to adding a linear classifier on top of the last layer of the model and  maximising the log-probability of the correct label \cite{Sun2019HowTF}. Here, we train 25 single-label classifiers ({\em i.e.}, one per ERC panel) and apply each of them to every single document, so that every single record can potentially belong from zero to N categories. Most of the documents in our final dataset actually belong to one category only, while some of them could not be labelled with any panel.


For constructing the training dataset, we use R\&I European projects funded with an ERC grant, available in CORDIS, and obtained from UNICS \cite{Gimenez2018UNiCSTO}. These projects are labelled with the ERC panel assigned by the competitive call. For each panel, we transform the project descriptions to vectors with SPECTER and compute the mass centre. Next, from the official portal for European data\footnote{\url{https://data.europa.eu/data/datasets/cordish2020projects?locale=en}}, we collect all scientific publications produced as a result of each ERC grant. Finally, we assign publications to the panel of their respective grant if their embeddings via SPECTER fall within a given distance from the centre of mass of the panel computed earlier. This allows us to build a train/test dataset composed of a mixture of project descriptions and publication abstracts.
For training, we randomly select up to 1,500 true publications and up to 20,000 negative publications per panel, while we test on project descriptions, as their panels are manually assigned, as we show in Table~\ref{dataset_size}.

\begin{table}[htttt]
\centering
\resizebox{0.95\textwidth}{!}{%
\begin{tabular}{l*{4}{c}r}
\hline
\multicolumn{1}{c}{} & \multicolumn{2}{c}{\textbf{Train \& Dev}} & \multicolumn{2}{c}{\textbf{Test}} \\
\cline{2-5}
\textbf{ERC Panel} & \textbf{True} & \textbf{False} & \textbf{True} & \textbf{False} \\
\hline
PE1 - Mathematics & 1500 & 14890 & 456 & 4640\\
PE2 - Fundamental Constituents of Matter & 1497 & 14972 & 554 & 4535\\
PE3 - Condensed Matter Physics & 1500 & 14948 & 498 & 4566\\
PE4 - Physical \& Analytical Chemical Sciences & 1500 & 14888 & 439 & 4607\\
PE5 - Synthetic Chemistry \& Materials & 1497 & 15029 & 565 & 4513\\
PE6 - Computer Science \& Informatics & 1497 & 15000 & 549 & 4511\\
PE7 - Systems \& Communication Engineering & 1440 & 14971 & 400 & 4579\\
PE8 - Products \& Processes Engineering & 1496 & 14943 & 536 & 4542\\
PE9 - Universe Sciences & 1500 & 14922 & 367 & 4596\\
PE10 - Earth System Science & 1497 & 14925 & 401 & 4603\\
LS1 - Molecules of Life: Biological Mechanisms, Structuresand Functions & 1500 & 14979 & 365 & 4552\\
LS2 - Integrative Biology: from Genes \& Genomes to Systems & 1497 & 14877 & 362 & 4686\\
LS3 - Cellular, Developmental \& Regenerative Biology & 1499 & 14879 & 326 & 4652\\
LS4 - Physiology in Health, Disease \& Ageing & 1499 & 14940 & 403 & 4567\\
LS5 - Neuroscience \& Disorders of the Nervous System & 1500 & 14973 & 486 & 4534\\
LS6 - Immunity, Infection \& Immunotherapy & 1499 & 14875 & 355 & 4649\\
LS7 - Prevention, Diagnosis \& Treatment of Human Diseases & 1500 & 14962 & 537 & 4578\\
LS8 - Environmental Biology, Ecologyand Evolution & 1500 & 14997 & 395 & 4549\\
LS9 - Biotechnology \& Biosystems Engineering & 1500 & 14833 & 290 & 4658\\
SH1 - Individuals, Markets \& Organisations & 829 & 14964 & 327 & 4578\\
SH2 - Institutions, Governance \& Legal Systems & 509 & 15405 & 471 & 4561\\
SH3 - The Social World \& Its Diversity & 1500 & 14902 & 306 & 4598\\
SH4 - The Human Mind \& Its Complexity & 1500 & 14950 & 452 & 4524\\
SH5 - Cultures \& Cultural Production & 639 & 15049 & 340 & 4479\\
SH6 - The Study of the Human Past & 1060 & 14975 & 384 & 4555\\
\hline
\end{tabular}
}
\addtolength{\tabcolsep}{-1pt}
\caption{Number of labels and examples in the train, dev and test sets of our dataset.}
\label{dataset_size}
\end{table}

To evaluate if noise reduction from data is necessary, we train on publications derived from ERC projects with their ERC categories, and we test on projects, whose labels are supervised. Table~\ref{result_domain_models} shows results for each classifier. Some of the domains are complex, and their borders are particularly fuzzy; however, the results are particularly satisfactory, even when we train and evaluate on different types of documents.

\begin{table}[htttt]
\centering
\resizebox{0.95\textwidth}{!}{%
\begin{tabular}{l*{4}{c}r}
\hline
\textbf{ERC Panel} & \textbf{P} & \textbf{R} & \textbf{F1} & \textbf{Acc.} \\
\hline
LS1 - Molecules of Life: Biological Mechanisms, Structures \& Functions & .886 & .854 & .869 & .966 \\
LS2 - Integrative Biology: from Genes \& Genomes to Systems & .811 & .815 & .813 & .95 \\
LS3 - Cellular, Developmental \& Regenerative Biology & .871 & .766 & .808 & .96 \\
LS4 - Physiology in Health, Disease \& Ageing & .846 & .894 & .868 & .959 \\
LS5 - Neuroscience \& Disorders of the Nervous System & .908 & .944 & .925 & .973 \\
LS6 - Immunity, Infection \& Immunotherapy & .885 & .942 & .911 & .975 \\
LS7 - Prevention, Diagnosis \& Treatment of Human Diseases & .832 & .829 & .831 & .937 \\
LS8 - Environmental Biology, Ecology \& Evolution & .886 & .962 & .92 & .975 \\
LS9 - Biotechnology \& Biosystems Engineering & .819 & .87 & .842 & .963 \\
PE1 - Mathematics & .986 & .935 & .959 & .987 \\
PE2 - Fundamental Constituents of Matter & .927 & .922 & .925 & .971 \\
PE3 - Condensed Matter Physics & .845 & .866 & .855 & .948 \\
PE4 - Physical \& Analytical Chemical Sciences & .853 & .834 & .843 & .952 \\
PE5 - Synthetic Chemistry \& Materials & .897 & .802 & .841 & .945 \\
PE6 - Computer Science \& Informatics & .925 & .903 & .914 & .968 \\
PE7 - Systems \& Communication Engineering & .904 & .78 & .829 & .958 \\
PE8 - Products \& Processes Engineering & .835 & .871 & .852 & .942 \\
PE9 - Universe Sciences & .986 & .964 & .974 & .993 \\
PE10 - Earth System Science & .941 & .944 & .943 & .983 \\
SH1 - Individuals, Markets \& Organisations & .974 & .842 & .896 & .978 \\
SH2 - Institutions, Governance \& Legal Systems & .842 & .906 & .871 & .953 \\
SH3 - The Social World \& Its Diversity & .779 & .794 & .787 & .949 \\
SH4 - The Human Mind \& Its Complexity & .906 & .945 & .924 & .974 \\
SH5 - Cultures \& Cultural Production & .922 & .821 & .864 & .969 \\
SH6 - The Study of the Human Past & .939 & .902 & .92 & .978 \\
\hline
\end{tabular}
}
\addtolength{\tabcolsep}{-1pt}
\caption{Evaluation metrics for the 25 classifiers, trained on publications and evaluated on projects. Trained by fine-tuning, fixing the following hyperpameter configuration: 4 epochs, learning rate of 2e-5, and batch size of 16. Input data is title+abstract. Results are reported as macro-average.}
\label{result_domain_models}
\end{table}

\section {Main Results}

We gathered, from a series of heterogeneous datasources, a dataset of scientific publications abstracts and R\&D projects descriptions for the entire STI ecosystem of Denmark, for the 2014--2019 time period (see Section \ref{sec:matMethods} for more details). We then tagged each single record by means of a controlled vocabulary for SDG 13 -- Climate Action (that is, we identified the vocabulary terms in each single text, by applying a series of textual matching rules). This enabled us to identify, within the initial dataset, all textual records linked with SDG 13. 

The number of documents that we could identify in each data source as well as those we could link with SDG 13 are reported in Table~\ref{results_records}. About 2\% of Scientific publications in Denmark between 2014 and 2019, both from OpenAire and OpenAlex, are related to our SDG of interest. In contrast, European projects are more linked, in relative terms, to the issue of Climate Action: this comes as no surprise,given the orientation of EU policies towards the sustainability issues.

\begin{table}[htt]
\centering
\caption{Number of records with at least one author affiliation or beneficiary from Denmark (2014-2019) and relative volume mapped to SDG 13.}

\addtolength{\tabcolsep}{1pt}
\begin{tabular}{lrr}
\hline
\textbf{Data source} & \textbf{Total Records in DK} & \textbf{Records related to SDG 13} \\
\hline
OpenAlex & 191,399 & 3,821 (2\%) \\
OpenAIRE & 235,906 & 5,273 (2.2\%) \\
CORDIS & 2,196 & 320 (14.6\%) \\
Kohesio & 294 & 14 (4.8\%) \\
\hline
\end{tabular}
\addtolength{\tabcolsep}{-1pt}
\label{results_records}
\end{table}

\begin{figure}[h]
\label{confetti}
\caption{t-SNE visualization of the 30 topics extracted from the SDG 13 corpus.}
  \centering
  \includegraphics[width=0.9\linewidth]{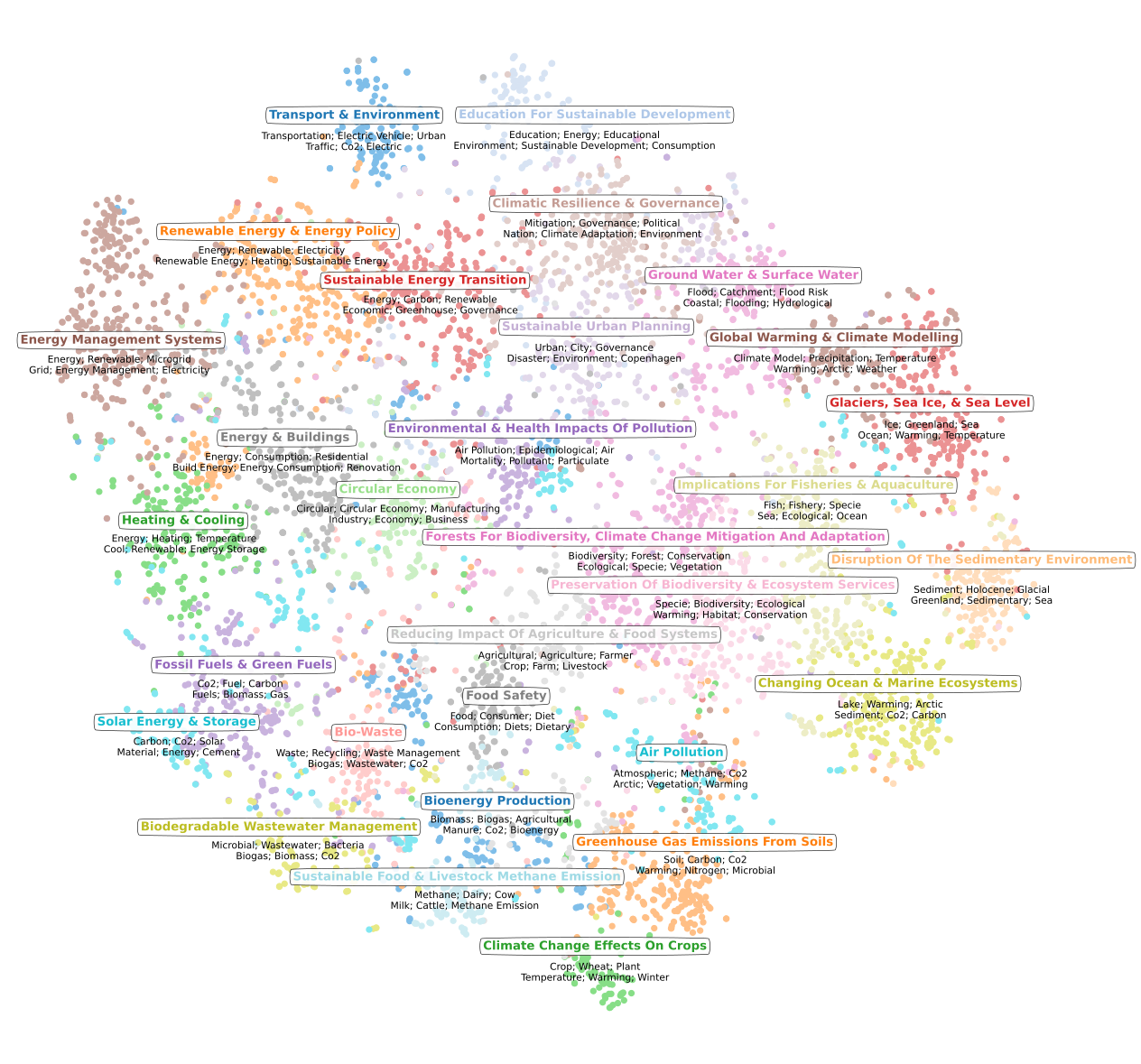}
  
\end{figure}

We finally applied both Topic Modelling and we classified per  ERC panels the SDG--related corpus, to obtain both a series of topics characterising Climate Action--related research and to gain a ``disciplinary'' view of such research. In Figure~\ref{confetti} we show a t-SNE visualisation of the automatically extracted topics \cite{van2008visualizing}, as we describe in Section \ref{sec:BERT}, from textual data in publications (from OpenAlex and OpenAIRE) and projects (from CORDIS and Kohesio) concerning SDG 13. 

In the graph, each dot is a single document: the clustering--based approach we follow in the present work allows us to link each record to a single topic. As one can observe, we were able to extract a series of thematically different topics from the SDG 13 corpus, each dealing with a different aspect of climate action. At the centre of the figure, one finds topics related with the environment, while going anti--clockwise from the top, it is possible to encounter topics related with energy, traditional and alternative fuels, emissions and pollution, impact on the biosphere and finally education and policy issues. A result like this may be extremely relevant for policy--makers, as it showcases what research niches are covered by the local STI ecosystem and it allow to inform decision--making.

Besides thematic information, policy--making is best informed via the identification of relevant actors in the identified STI landscape. In this case, the identification of the most relevant actors across different data sources is affected by the different degree of affiliation disambiguation inside each individual source. Here we analyse the most active actors for research on SDG 13 per source is shown in Table~\ref{tab:top_aff_sdg13}, 
as they emerge from each data source. Indeed, the same actor can appear with several name variations in affiliations reported on papers. Different repositories have a different degree of disambiguation of the affiliations. 

\begin{table}[h]
\caption{Top-ten affiliations per each dataset in SDG 13-related documents.}

\label{tab:top_aff_sdg13}
\begin{center}
\scriptsize
\resizebox{\textwidth}{!}{%
\begin{tabular}{p{3cm}p{0.8cm}|p{3cm}p{0.8cm}|p{3.8cm}p{0.8cm}|p{3cm}p{0.8cm}}

\hline
\textbf{OpenAlex}& \textbf{\#SDG13 } &\textbf{OpenAIRE}& \textbf{\#SDG13 } &\textbf{CORDIS}& \textbf{\#SDG13 } &\textbf{Kohesio}& \textbf{\#SDG13 }\\
\hline
University of Copenhagen&1,016&Technical University of Denmark&1,589&DANMARKS TEKNISKE UNIVERSITET&66&Den Erhvervsdrivende Fond Development Centre UMT&2\\
Aarhus University&955&Aarhus University&1,053&KOBENHAVNS UNIVERSITET&59&Vejle Kommune&2\\
Technical University of Denmark&765&Aalborg University&954&AARHUS UNIVERSITET&43&Kolding Municipality&2\\
Aalborg University&433&IT University of Copenhagen&877&AALBORG UNIVERSITET&26&Fredericia Municipality&2\\
University of Southern Denmark&203&University of Copenhagen&506&TEKNOLOGISK INSTITUT&10&CLEAN&1\\
Geological Survey of Denmark and Greenland&137&University of Southern Denmark&379&DANMARKS METEOROLOGISKE INSTITUT&10&Aalborg Municipality&1\\
Danish Meteorological Institute&119&KOBENHAVNS UNIVERSITET&191&Geological Survey of Denmark and Greenland&7&Aalborg University&1\\
Roskilde University&43&United Nations Office for Project Services&116&INNOVATIONSFONDEN&6&Aarhus Municipality&1\\
Copenhagen Business School&40&Roskilde University&111&ENERGISTYRELSEN&5&Odense Kommune&1\\
Aalborg University – Copenhagen&34&DCE - Danish Centre for Environment and Energy&48&COPENHAGEN BUSINESS SCHOOL&4&Sidis Robotics ApS&1\\
\hline
\end{tabular}
}
\end{center}
\end{table}

In the OpenAlex dataset, the actor with the most records is the University of Copenhagen, with more than 55,000 documents, which account for about 30\% of all Denmark publications. (A few hundred documents are instead associated with faculties of the same university). 1.84\% of these publications are in SDG 13.  Aarhus University (35,956; 2.66\%), Technical University of Denmark (23,368; 3.27\%)  and Aalborg University (20,844; 2.08\%) are the other most productive universities. The main Danish institutional actors, in terms of contribution to research aligned with SGD 13,  have been identified. Besides more or less different numbers and shares found for the large universities -- University of Copenhagen, Aarhus University, Aalborg University and Technical University of Denmark --  actors with high shares across all data sources are for example Geological Survey of Denmark and Greenland and the Danish Meteorological Institute are observed across all datasets (except in Kohesio, where actors are either municipalities or universities). However, it is possible to observe cases like Roskilde University, that appears to have double the share of SDG 13 documents in OpenAIRE compared to OpenAlex (2.64\% against 1.34\%).

Unfortunately, as one can clearly see, the limited interoperability across data sources greatly limits the type and depth of the analyses that can be carried out directly by leveraging the data as it comes from the different data sources. A lengthy (and sometimes painstaking) exercise of name cleaning and reconciliation must be carried out to provide clear cut analyses that could allow decision makers to design targeted policies.

As a final pilot exercise, we proceeded to classify the documents linked to SDG 13 per ERC panels. To do so, we trained a Deep Learning textual classifier on a weakly supervised dataset (see Section \ref{sec:matMethods}) and we applied it to our Danish SDG 13 corpus. This effort allowed us to obtain a disciplinary classification of the records which is consistent across data sources and which may enable, in turn, a comparison of the Danish STI ecosystem with other geographical perimeters of interest.

\begin{figure}[h]
\caption{Number of documents per source and by ERC Panel.}
\label{number_erc_source}
  \centering
  \includegraphics[width=0.95\linewidth]{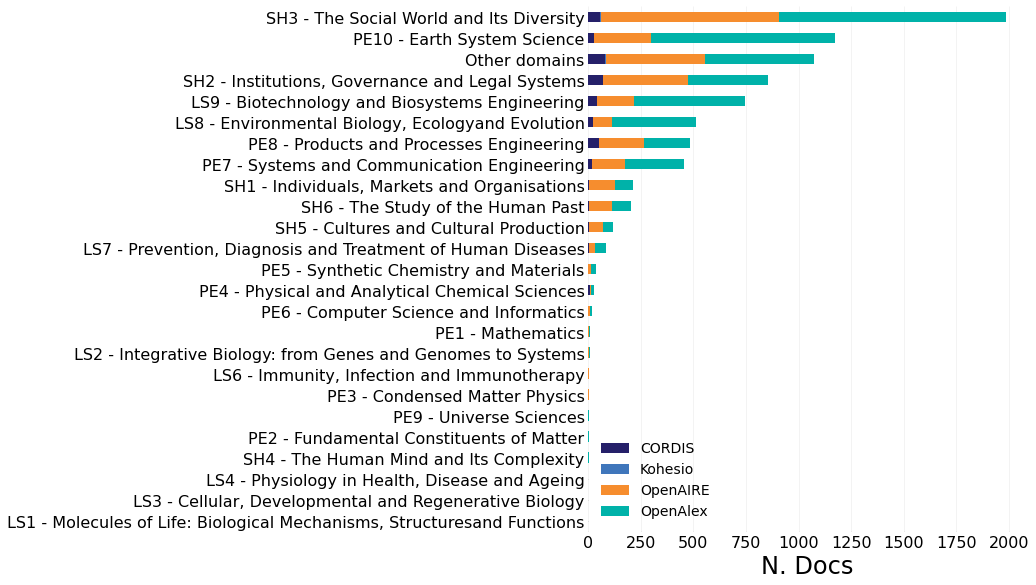}
  
\end{figure}

In Figure~\ref{number_erc_source} we present distribution of documents by source and ERC panel. Perhaps surprisingly, the majority of the STI documents analysed was linked to Social Science issues, followed by Earth Sciences. Also interestingly (and underscoring the importance of cross--platform analyses such as this one), one can see that the various data sources have a different coverage of the panels. Although our methodology may potentially link each single document to more than one panel, the majority of records were actually assigned one ERC panel only: however, in Fig. \ref{number_erc_source} some multiple counting occurs, as it normally happens with bibliometric categories.

\begin{figure}[h!t]
  \caption{Co-occurrence matrix between topics and ERC panels. The labels for the topics in x-axis are: {\scriptsize 0 = Sustainable food \& Livestock Methane Emission; 1 = Biodegradable wastewater management; 2 = Climate change effects on crops; 3 = Implications for fisheries \& aquaculture; 4 = Transport \& Environment; 5 = Sustainable Energy Transition; 6 = Renewable Energy \& Energy Policy; 7 = Climatic Resilience \& Governance; 8 = Education for sustainable development; 9 = Sustainable urban planning; 10 = Solar Energy \& Storage; 11 = Energy Management systems; 12 = Global Warming \& Climate Modelling; 13 = Glaciers, Sea Ice, \& Sea Level; 14 = Air Pollution; 15 = Disruption of the sedimentary environment; 16 = Preservation of Biodiversity \& Ecosystem Services; 17 = Fossil Fuels \& Green Fuels; 18 = Changing Ocean \& Marine Ecosystems; 19 = Reducing impact of agriculture \& food systems; 20 = Bio-waste; 21 = Food safety; 22 = Forests for biodiversity, climate change mitigation and adaptation; 23 = Environmental \& Health Impacts of Pollution; 24 = Bioenergy production; 25 = Heating \& Cooling; 26 = Greenhouse gas emissions from soils; 27 = Ground Water \& Surface Water; 28 = Energy \& Buildings ; 29 = Circular Economy}}
\label{heatmap}
  \centering
  \includegraphics[width=\linewidth]{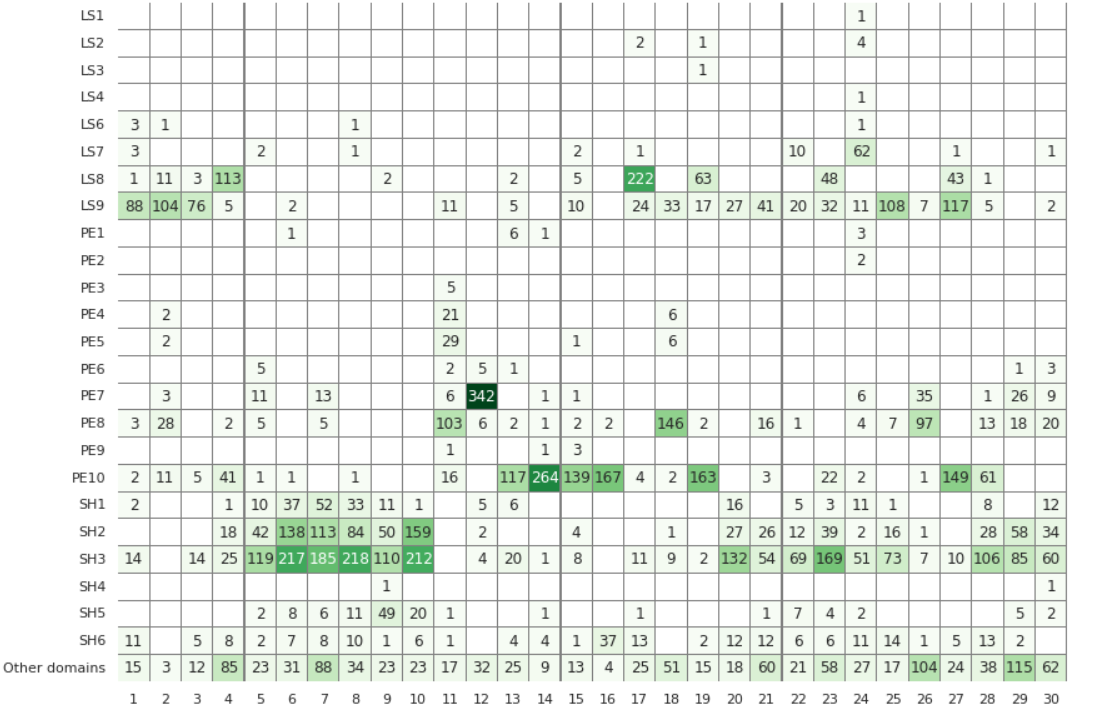}

\end{figure}

As a cross comparison between the topic approach and the ERC panel focus, we show in Figure~\ref{heatmap} the co-occurrence matrix between ERC panels and topics obtained via topic modelling. By means of this mapping, we observe a strong correspondence between the topic ``Global Warming \& Climate Modelling'' and the Panel Physics \& Engineering 7 (PE7), Systems \& Communication Engineering and then an interesting cluster of topics about the transition towards sustainable energies and panels Social Sciences and Humanities 1, 2 and 3 (SH1--3), suggesting a considerable interest of the Danish STI ecosystem towards the issues of policy regulation for energetic transition. In contrast, a very little volume of Danish research on Climate Action could be linked to the Life Sciences Panels (LS). Notably, a great volume of STI outputs could not be linked to any ERC panel. This is especially the case for the topic ``Greenhouse Gas Emissions from Soils'', which is equally split between the ERC panel ``Product \& Process Engineering'' and no panel at all, and the topic ``Circular Economy'', which could be mapped back to a couple of Social Sciences panels, but for which a great fraction of records could not be linked back to any ERC panel.

\section{Conclusions and further work}
In this paper, we presented a proof--of--concept study of the use of Open Resources to map the research landscape on SDG 13 -- Climate Action, for an entire country, Denmark. This type of mapping exercise is extremely useful for STI decision--makers, who, to design effective policies within their respective sphere of influence, need to have a clear vision of {\em what} is researched and by {\em whom}.

Here, we carried out a study of this sort by relying on Open Data for Research Projects (gathered from CORDIS and the Kohesio platform) and Scientific publications (collected from Open Aire and OpenAlex), by using an open vocabulary for mapping STI records on SDG 13 and by using openly available Deep Learning models to classify the corpus in accordance with the 25 ERC panels. The results we obtain are fairly encouraging: the coverage of the data analysed is extensive, both in absolute terms\footnote{Although we make no mention in the text, we made an overall comparison with a ``traditional'' bibliometric database such as Scopus: the number of records that could be retrieved from Scopus, for the same country, in the same time period, was lower than what found both in OpenAire and OpenAlex} and in terms of scientific disciplines and actors. Interestingly, the data sources analysed offer a complementary view of the research domains, and allow one, when used in combination, to obtain a wide and precise overview of the local STI ecosystem.

This study also identified some room for improvement: the lack of interoperability across data sources (notably, the absence of a common standard for identifying organisations) and a limited degree of data curation and reconciliation ({\em e.g.}, identical affiliations appearing with different names within the same data source) hinder the possibility of applying these data sources for analyses of this sort in a straightforward manner. Here, we presented the data as returned from the different sources, but a real application of a work of this sort would require a lengthy and costly data cleaning and alignment process. On the positive side, the wide availability of textual data offered by the sources employed in this study enables to make very flexible custom analyses and to overcome at least the taxonomic non--interoperability across sources.

The main findings of this study are that Open Science initiatives can be a formidable resource for the sake of informing policies via STI ecosystems mapping. Open Scholarly platforms can be especially useful when different resources, covering different types of outputs and disciplines, are used in combination. Nevertheless, current initiatives (albeit encouraging from the point of view of coverage) are still very limited in terms of interoperability, thus hindering the immense potential they all have. The principal output of this study is therefore a call to:

\begin{enumerate}
    \item continue advancing in the creation and use of open data standards and platforms;
    \item better define standards to allow interoperability across initiatives;
    \item explore policy frameworks and methodologies to help connect STI actors and activities with transformative results

\end{enumerate}

Several next steps may be explored starting from this proof--of--concept: first of all, further STI sources can be integrated in the study, to capture even other actors and value chain segments of local STI ecosystem. For example, patent data could be integrated in the study, so to capture industrial innovation contributions. Secondly, further data on, say, R\&D projects could be collected by other parties, such as national research and philanthropic funders. For the case of Denmark, specifically, the weight of the Novo Nordisk Foundation in funding research has a considerable impact in steering research at the national level. Finally, further scientific domains (other than the SDGs) and disciplinary categories (other than ERC panels) can be explored by using the same methodology we adopted in this study. We leave all those potential further avenues of research to future studies.

\section*{Acknowledgments}
This work was partly funded by the European Commission H2020 Programme via the INODE project, under grant agreement No 863410.

\bibliography{bibliography}
\end{document}